# Electro-optical silicon isolator

Hugo Lira[1], Zongfu Yu[2], Shanhui Fan[2], Michal Lipson[3*]

[1] *School of Electrical and Computer Engineering, Cornell University, Ithaca, NY 14853, USA*
[2] *Department of Electrical Engineering, Stanford University, Stanford, California 94305, USA*
[3] *Kavli Institute at Cornell for Nanoscale Science, Cornell University, Ithaca, NY 14853, USA*
*To whom the correspondence should be addressed. Email: michal.lipson@cornell.edu

**Optical isolators are essential components in optical networks and are used to eliminate parasite reflections that are detrimental to the stability of the optical systems. The challenge in realizing optical isolators lies in the fact that in standard optoelelectronic materials, including most semiconductors and metals, Maxwell's equations, which governs the propagation of light, are constraint by reciprocity or time-reversal symmetry. As a result, standard optical devices on chip, including some of the passive metallo-dielectric structures recently explored for isolation purposes[1-2], all have a *symmetric* scattering matrix and therefore fundamentally cannot function as an optical isolator. In order to break this symmetry, traditional optical isolators rely upon magneto-optical effects [3-8], which require materials that are difficult to integrate in current micro-electronic systems. The breaking of this symmetry without the need for magnetic materials has been a long-term goal in photonics. Nonlinear structures for optical isolation have been explored [9-12]. However, these typically provide optical isolation only within a particular range of operating power. Here we create a non-magnetic CMOS-compatible optical isolator on a silicon chip. The isolator is based on indirect interband photonic transition, induced by electrically-driven dynamic refractive index modulation. We demonstrate an electrically-induced non-reciprocity: the transmission coefficients between two single-mode waveguides become dependent on the propagation directions only in the presence of the electrical drive. The contrast ratio between**

**forward and backward directions exceeds 30dB in simulations. We experimentally observe a strong contrast (up to 3 dB) limited only by our electrical setup, for a continuous-wave (CW) optical signal. Importantly, the device is linear with respect to signal light. The observed contrast ratio is independent of the timing, the format, the amplitude and the phase of the input signal.**

Indirect interband photonic transitions can enable optical isolation: Indirect transitions fundamentally break the time-reversal symmetry and the reciprocity relation that connects the forward and backward propagating light, and interband transitions provide a spectrally-pure output that is free from modulation side bands. Both of these aspects are essential for successful operation of an optical isolator[13]. Photonic transitions in highly confined optical structures, where modes can be carefully manipulated and tailored, are of fundamental interests due to conceptual analogy with electronic transitions in semiconductors[14]. An indirect interband photonic transition occurs between two optical modes having different longitudinal wavevectors (hence the word "indirect"), and having transverse modal profiles with different symmetries (hence the word "interband" since the two modes belong to different photonic bands). Up to now, all observed photonic transitions, including the recent observation of photonic transition in micro-ring resonators[15], and photonic transitions in conventional travelling-wave electro-absorption modulators, are intraband transition between two modes with the same transverse modal profiles. The intraband transition generates an output of a significant frequency comb that is difficult to use in an optical isolator. Demonstrating indirect interband transition is more challenging. In a waveguide system, to induce an interband transition between two modes with different symmetry in its transverse modal profile, the modulation itself cannot be uniform across the waveguide cross-section. In addition, since the typical frequency of high speed modulators[15] is on the order of few GHz, the wavevector difference, between two different transverse modes having frequencies separated by a few

GHz, is typically quite large and hence the modulation needs to be specifically constructed. Conventional travelling-wave electro-optic modulator design cannot generate such a large required modulation wavevector[15].

Here we demonstrate indirect interband photonic transitions by tailoring the dispersion of a silicon photonic structure to match the frequency and wavevector of a specifically designed traveling electrical wave modulation. The structure used is a slotted waveguide shown in Fig. 1a that is engineered to have two optical modes (even and odd). The dispersion of the structure shown in Fig. 1a and b is tailored so that the two modes are separated in frequency by only a few GHz, for a difference in wavector $k=2\pi/\lambda$ where $\lambda$ is on the order of only a few hundreds of microns. This enables one to create transition between the two modes by applying an electrical signal with a GHz frequency that is achievable in silicon electro-optic devices[16], and to use a structure that is compact ( on the order of a few hundred microns in length). Isolation based on such structure is achieved, since the photonic transition is not allowed when light propagates in the forward direction from left to right, and allowed when light propagates in the opposite backward direction (Fig. 1 b left). This is because at the frequency of modulation, in the forward direction, $|k_{even}-k_{odd}| \neq |k|$, while in the backward direction, $|k_{even}-k_{odd}|_{RL} =|k|$ (see Fig. 1 b ). Figure 1c shows the mode amplitudes in each direction obtained from our simulation using coupled mode equations [13] of a continuous wave input into the structure assuming 10 GHz index modulation in the structure with dimensions $w$=450nm, $h$=215nm, $t$=35nm and $d$=500nm. One sees total conversion in the backward direction and minimal conversion (<2% in the example shown) in the forward direction. We emphasize the CW nature of the simulation for both the modulation and the optical input. The isolation has nothing to do with the relative timing between the optical signal and the modulation.

The isolator based on such indirect transition is achieved by using 1x2 MMI (Multi-Mode Interference waveguide) on both ends of the slotted waveguide (Fig. 2a). The MMI serves to couple light to and from a single mode waveguide (with an even mode) to the even mode of the slotted waveguide, and to filter the converted odd modes propagating in the slotted waveguide propagating in the backward direction. In the forward direction the photonic transition is not allowed and thus the even mode of the single-mode input waveguide couples back to the output waveguide through the MMI at the end of isolator. In contrast, in the backward direction (right to left), the transition is allowed and therefore the even mode is completely converted to the odd mode, which is filtered by the MMI and cannot couple to the output single-mode waveguide. As a result, the odd mode is dissipated as radiation. Fig. 2b shows a 2D time-domain FEM (Finite Element Method) simulation of the isolator. Note that in order to induce conversion, the overlap integral between the modulation spatial distribution and the initial and final modes should be non zero [13] and therefore in the simulation the modulation is applied only to the lower half of the two-mode waveguide. The input signal was set at a wavelength of 1550 nm, and the permittivity modulation has a maximum $\Delta\varepsilon=0.23211$, at 10 THz with wavenumber $q=541391$ m$^{-1}$. The total length of the waveguide is 30μm. Also, simulations with modulation frequency reduced down to 100 GHz were successfully performed, with $q=25984.8$m$^{-1}$, $\Delta\varepsilon=0.0182$ and device length of 0.413 mm, with mode conversion observed in only one direction, which shows the scalability of the principle to frequencies that can be achieved in current electrical diodes on optical waveguides [17-18]. Importantly, the difference in the transmission coefficients between two single-mode waveguides directly proves that the system has a non-symmetric scattering matrix.

The electrical traveling wave modulation that induces the indirect photonic transition is realized here by embedding the slotted waveguide within two transmission lines connected to the waveguide via electrical diodes. The modulation of the index is achieved by creating depletion regions along the waveguide. We use a modulation frequency of 10GHz. At this modulation frequency, the electrical wave along the transmission line has a large period above a centimeter and a corresponding small wavevector, which is not suitable for achieving the interband transition that we require here. To achieve a larger modulation wavevector, we design the structure in which modulation wavevector is instead encoded in the pn-junction configurations. We discretize each spatial period of the required index modulation into four separate regions. In each period, a single (alternating) applied voltage to the device then induces four different depletion regions with different length, i.e. different carrier concentrations (and therefore index change) along the slotted waveguide. These four different concentrations are achieved by using two different junctions (pn-np and np-pn) placed in alternating positions along the waveguide and by using two transmission-lines with a $\pi/2$ phase difference in their voltages as shown in Fig. 3a and b. Under an applied bias, only the reverse-biased diodes experience carrier depletion inducing index change while the forward-biased diodes experience only minimal carrier leakage. In addition when a large (small) voltage $V$ is applied, the diode connected to one transmission line induces a large (small) index change in the waveguide while the one connected to the other transmission induces a small (large) index change in the waveguide. Fig 3b shows the diodes that experience a reverse-bias voltage at a given moment, represented by the colored diodes (green, blue, red and gray). Such arrangement only induces index modulation in part of the cross section of the waveguide, fulfilling the required non-zero overlap condition. We choose to operate at a modulation frequency of 10 GHz, which corresponds to the difference in frequency between the two modes for the waveguide shown in Fig. 1b above. The achieved modulation of the refractive index along the waveguide is shown in Fig. 3c. Each dot corresponds to a diode along the propagation axis. The gray and red dots correspond to pn-np and np-pn diodes connected to one transmission line, while the blue and green correspond to the pn-np and np-pn diodes in the the second transmission line having a $\pi/2$ phase difference. The achieved fundamental harmonic of the index modulation is shown in orange. One can clearly see that each modulation period $\lambda_m$ of 450 µm is discretized in 4 parts, with period slightly reduced due to the traveling wave voltage distribution across the transmission line (green and gray dotted lines in Fig. 3c). Each diode is designed to be 110-μm long separated by a 2.5-μm region with the opposite dopant providing electrical insulation. The overall number of discrete modulation periods is 22, i.e. 88 modulation sections, or 166 diodes. In order to prevent the periodically-loaded transmission lines from having a cut-off below the desired modulation frequency we add spiral inductors in parallel with each of the pn-np and np-pn junctions in the waveguide (with a total length 1.5 mm corresponding to an $L \approx$ 1.84 nH). The length of the stub connecting the capacitors and inductors to the transmission line also affects the cut-off frequency, and from design we expect that a 100-μm long stub would push the cut-off above 10 GHz. In order to induce maximum index modulation with minimal loss the n-doped regions and p-doped regions concentrations were chosen to be $1 \times 10^{18} cm^{-3}$ $1 \times 10^{17} cm^{-3}$ respectively, with the center of the dopant region shifted about 190 nm from the center of each of the waveguides, so that losses are minimized for the index change required of the device.

We measure the forward and backward transmission spectra by inputting a continuous wave optical signal, applying a 10Ghz modulation and swapping the input and output fibers. Note that the amplitude of the output signal is not affected by the modulation since only the phase (i.e. refractive index) is modulated. Fig, 4 shows the ratio between the two spectra. When no electrical signal is applied the transmission is completely reciprocal with a unity contrast ratio between the forward and backward direction. When the electrical signal is applied a clear dip appears in the spectrum, indicating nonreciprocity induced by photonic transition, as shown in Fig. 4. The spectral width of the dip corresponds to the bandwidth of the device and is determined by the wavelength dependence of the difference in *k*-vector of the even and odd modes. As we continue to increase the power of the applied modulation (i.e. stronger coupling between the even and odd modes), we obtain greater contrast between forward and backward transmission. We measured up to 3 dB isolation when operating at a wavelength of 1558 nm. This isolation is smaller than the one obtained from simulation due to limitations with our electrical signal power supply, which can achieve a maximum 25 dBm output power (or 5.6 $V_p$ applied, which might be smaller due to reflections caused by impedance mismatch). For comparison we show in the bottom right of Fig. 4 the simulated relative transmission using the mode conversion equations and considering the dispersion of the waveguide to determine the bandwidth. One can see that the results agree well with the expected performance of the device.

The isolator shown here is key for future photonic systems on chip. The isolation degree, insertion loss, bandwidth and power efficiency of the device in principle could be increased using appropriate waveguide and electrical elements design and fabrication. The isolation degree can be increased by having better impedance match and higher input power for the electrical signal. The insertion loss can be reduced by designing the pn junctions. The dopants in the diodes are expected to induce loss on the order of 16 dB for a 1.0 cm device. This loss could be decreased greatly by using alternative pn-junction schemes such a recently demonstrated in[19], which also reduce the power consumption and increase the isolation attained. The bandwidth of operation can be increased by tailoring the dispersion of the structure so that photonic bands of the even and odd modes are parallel to each other. The bandwidth depends on how parallel the dispersion relations for the even and odd waveguides, and can be engineered by changing dimensions and geometry of the waveguides.

**Acknowledgement**

This work was performed in part at the Cornell NanoScale Facility, a member of the National Nanotechnology Infrastructure Network, which is supported by the National Science Foundation. This work was supported in part by the NSF through CIAN ERC under Grant EEC-0812072. Hugo Lira thanks his sponsorship support provided by the Brazilian Defense Ministry and insightful discussions with Prof. Ehsan Afshari. The authors also acknowledge the support in part of the AFOSR-MURI program (Grant No. FA9550-09-1-0704).

**Methods**

We fabricate the device on a SOI platform in a completely CMOS-compatible process. The PMMA photoresist masks for the dopants are written using e-beam lithography, followed by implantation of $B^+$ with a concentration of $1\times10^{17} cm^{-3}$ and the $P^-$ with a concentration of $1\times10^{18} cm^{-3}$. Masks for highly doped regions are written as well, followed by implantation of $1\times10^{20} cm^{-3}$ of $BF_2^+$ and $Ar^-$ to form low resistance region for accessing the p and n regions, respectively, for electrical contacts. Next we write a maN-2403 photoresist mask with the waveguide pattern using e-beam lithography, and etched the silicon down 215 nm, leaving a thin 35-nm slab everywhere. The dopants are then activated on anneal furnace and RTA process, followed by cladding the waveguides with a 1-µm thick $SiO_2$ deposited with PECVD tool. We then write the mask for vias and inductors, etching through the $SiO_2$ and sputtering 100 nm of $MoSi_2$ for low resistance contacts and 600 nm of AlCuSi. Another cladding layer is deposited, 600-nm thick, and a second set of vias are etched. Finally, we write the mask for lift-off of a 1500 nm deposition of AlCuSi to fill up the vias and form the transmission lines. In the left side of Fig. 4 we show an image of the electrical elements of the isolator (top) and of the optical elements (bottom) of the fabricated device. The insets point out to the inductors we place in parallel to the diodes, and to the vias which contact the pn-np diodes.

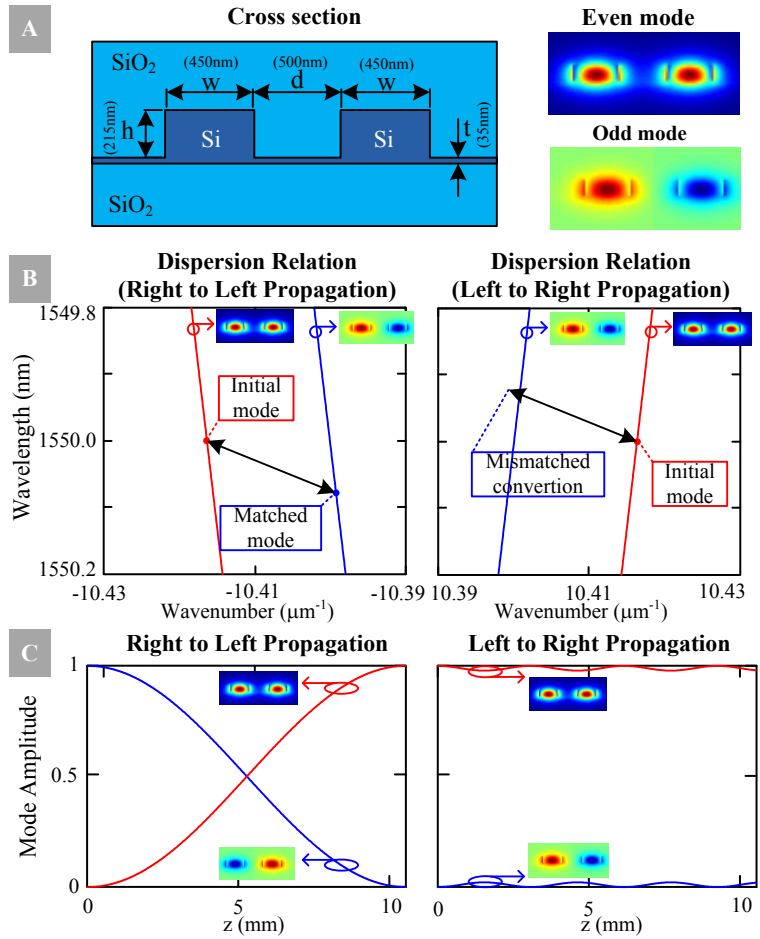

Fig. 1: (a) Waveguide geometry and materials. (b) Dispersion relation for even (red) and odd (blue) modes of coupled waveguides. The black arrow represents the traveling-wave index modulation. In the left, it matches the initial mode to another mode, while in the right it is observed conversion mismatch. (c) Dynamics of the mode conversion. Right to left propagation achieves full conversion from one mode to the other, while left to right propagation does not.

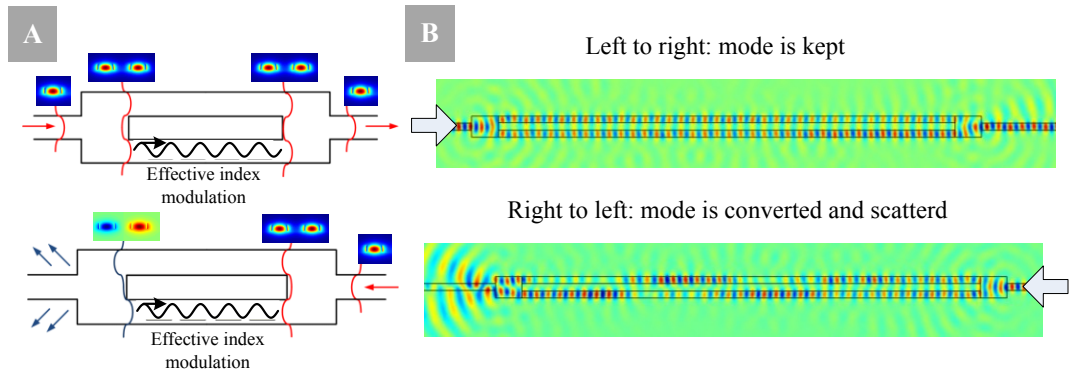

Fig. 2: (a) Schematic of the isolator. A single mode waveguide feeds a 1x2 MMI, which provides the even mode for the isolator. By modulating the refractive index of one waveguide we obtain a non-zero overlap between the modes and modulation. In one direction the even mode is converted to the odd mode, but it is not converted in the other direction. (b) FEM time-domain simulation showing the conversion occurring in only one propagation direction.

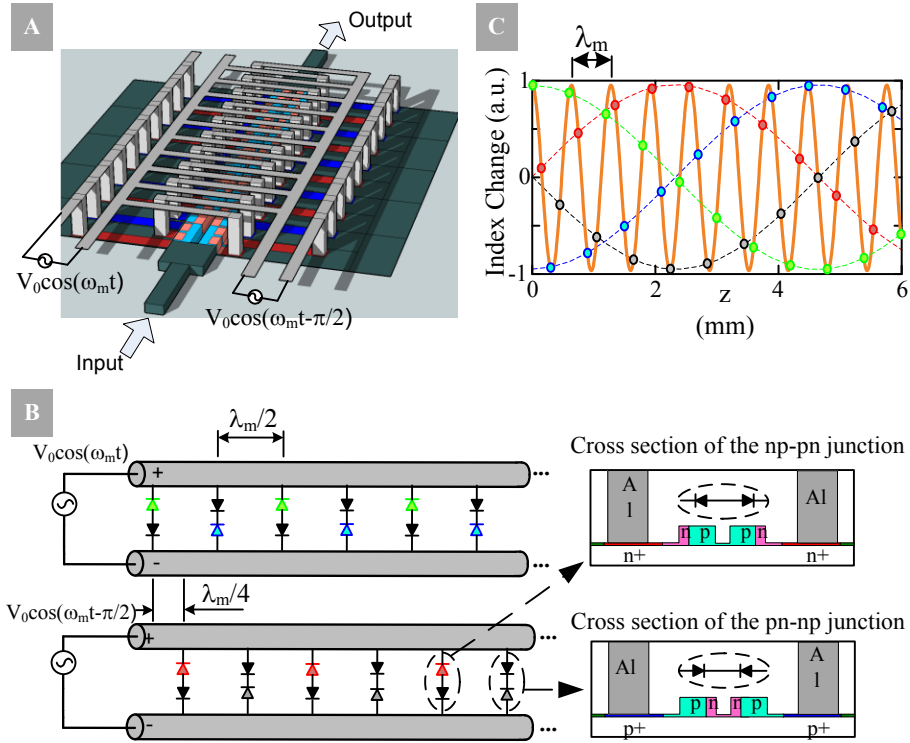

Fig. 3: (a) Simplified schematics of the device. Undoped silicon is shown in green, lightly p-doped silicon in light blue, lightly n-doped silicon in rose, heavily p-doped silicon in dark blue, heavily n-doped silicon in red, vias and electrical wiring in gray. (b) Schematics of the two transmission lines feeding pn-np and np-pn junctions. Reverse biased diodes are represented in green, blue, red and gray while forward biased diodes are in black. The insets shown the distribution of dopants across the waveguide forming np-pn and pn-np junctions. (c) Normalized index change. Each dot represent a diode in reverse bias along the waveguide. Green and blue dots are pn-np and np-pn diodes fed by one transmission line, while red and gray are pn-np and np-pn diodes fed by the delayed transmission line. In orange we have the fundamental harmonic of the discretized modulation achieved.

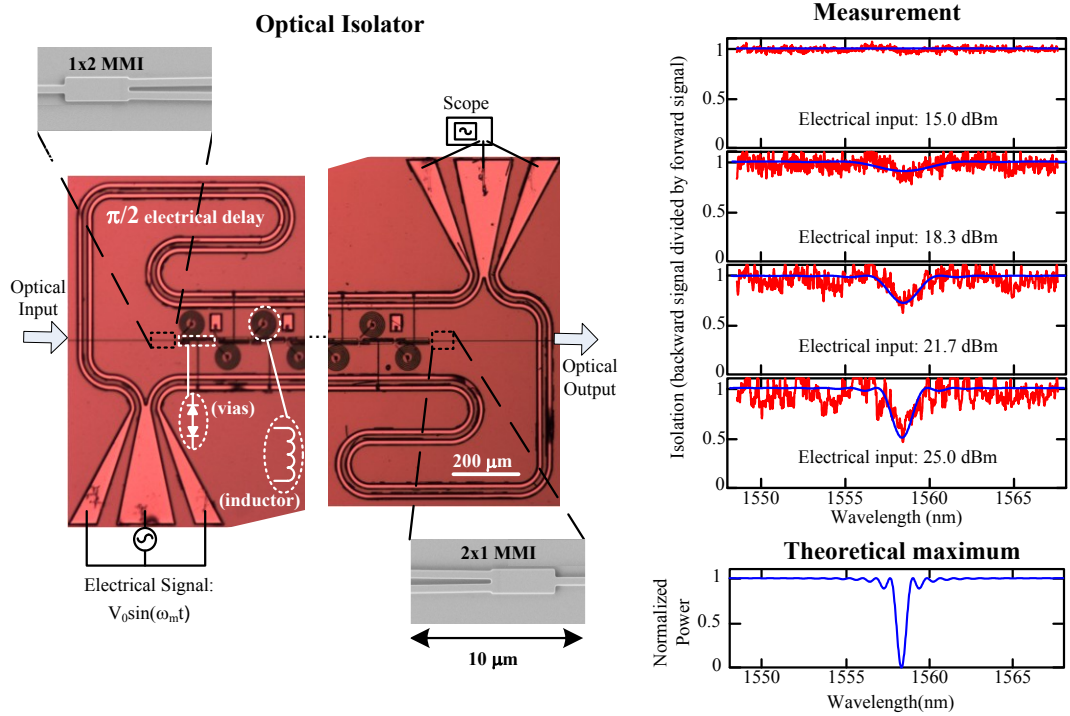

Fig. 4: In the left we show the electrical elements (Optical Microscope) and optical elements (Scanning Electron Microscopy) of the electrically-driven optical isolator. In the right we show sequentially the increase of isolation measured as a function of electrical signal input power (red). We observe up to 3 dB isolation with a electrical input of 25 dBm, and smaller values as the input decreases. The blue line is the fitting considering the conversion equations[13]. In the bottom right we show the maximum isolation expected and the bandwidth of conversion for the dispersion extracted from the fitting of the measured data.